\documentclass[usenatbib]{mn2e}
\usepackage[dvips]{color}
\usepackage{graphicx}
\usepackage{lscape}

\title[Discovery of a Cyclotron Absorption Line in RX J0440.9+4431]
{Broadband Observations of the Be/X-ray Binary Pulsar RX J0440.9+4431: Discovery of a Cyclotron Absorption Line}
\author[Tsygankov, Krivonos, Lutovinov]{S.\,S.\,Tsygankov$^{1,2,3,4}$\thanks{E-mail:
tsygankov@iki.rssi.ru}, R.\,A.\,Krivonos$^{3,4}$, A.\,A.\,Lutovinov$^{4,2}$\\
$^{1}$Finnish Centre for Astronomy with ESO (FINCA), University of Turku,  V\"ais\"al\"antie 20, FI-21500 Piikki\"o, Finland \\
$^{2}$Astronomy Division, Department of Physics, FI-90014 University of Oulu, Finland\\
$^{3}$MPI for Astrophysics, Karl-Schwarzschild Str. 1, Garching, 85741, Germany\\
$^{4}$Space Research Institute of the Russian Academy of Sciences, Profsoyuznaya Str. 84/32, Moscow
  117997, Russia}
\begin{document}

\date{Accepted 2011 December 30}

\pagerange{\pageref{firstpage}--\pageref{lastpage}} \pubyear{2011}

\maketitle

\label{firstpage}

\begin{abstract}
{We report the results of an analysis of data obtained with the INTEGRAL,
Swift and RXTE observatories during the 2010 April and September outbursts
of the X-ray pulsar RX\,J0440.9+4431. The temporal and spectral properties
of the pulsar in a wide energy band ($0.6-120$ keV) were studied for the
first time. We discovered a $\sim32$ keV cyclotron resonant scattering
feature in the source spectrum, that allowed us to estimate the magnetic
field strength of the neutron star as $B\simeq3.2\times10^{12}$ G. The
estimate of the magnetic field strength was confirmed by a comprehensive
analysis of the noise power spectrum of the source. Based on the recurrence
time between Type I outbursts the orbital period of the binary system can be
estimated as $\sim155$ days. We have shown that the pulse profile has a
sinusoidal-like single-peaked shape and has practically no dependence on the
source luminosity or energy band.}

\end{abstract}

\begin{keywords}
X-ray:binaries -- (stars:)pulsars:individual -- RX\,J0440.9+4431
\end{keywords}

\section{Introduction}

It is conventionally believed that typical X-ray binary pulsars with Be
optical companions (hereafter Be/XRP) manifest themselves through transient
activity of two types (see, e.g., \cite{reig2011}):

- Type I outbursts are caused by the increase of the mass accretion rate
onto the neutron star during the periastron passage. Such flares are usually
characterized by their periodic appearance (once per binary orbit), short
duration (small part of the orbital period) and peak X-ray luminosity about
or less than $\sim10^{37}$ erg s$^{-1}$;

- Type II outbursts deal with the non-stationary increase of an amount of
matter in the circumstellar disc around the Be star and can occur at any
orbital phase. The duration of such events varies from weeks to months,
during which the source X-ray luminosity can reach the Eddington limit
($\sim10^{38}$ erg s$^{-1}$).

\cite{reig1999} pointed out the existence of a subclass of binaries with Be
companions which harbor a slowly rotating neutron star and are characterized
by a persistent low-luminosity X-ray emission. Those authors proposed also
the model for such sources in which the neutron star, orbiting around a Be
star in a relatively wide binary system, accretes material only from the
low-density outer regions of the normal star envelope. Studies of such
long-period low-luminosity sources are crucial to understand the Be/XRP
binary systems formation processes and investigations of the interaction
between the Be star circumstellar disc and the neutron star.

The X-ray source RX\,J0440.9+4431 was found during the ROSAT Galactic plane
survey with the optical companion {\rm BSD\,24-491/LS\,V\,+44\,17}
classified as a Be star \citep{motch97}. Later \cite{reig1999} detected
X-ray pulsations from this source with the period of $\sim202.5$ s and
related it to the not numerous family of persistent Be/XRP binaries with
neutron stars. Based on the optical observations the distance to the system
was estimated as $3.3\pm0.5$ kpc \citep{reig2005}.

The first evidence of the pulsar outburst activity in X-rays was found by
\cite{morii2010} in late March 2010 with the MAXI all-sky monitor. The next
outburst was repeated after $\sim5$ months of a quiescence state in
September 2010 \citep{kri2010a}. During the latter flare the pulsar was in
the INTEGRAL observatory field of view, that gave us the opportunity to
investigate the properties of RX\,J0440.9+4431 in hard X-rays for the first
time. The third outburst detected by the Swift observatory in late January
2011 allowed \citet{tsy2011} to propose the existence of the $\sim155$-days
orbital period in the binary system. During these outbursts the X-ray
luminosity of the source reached $\sim9\times10^{36}$, $\sim3\times10^{36}$
and $\sim2\times10^{36}$ erg s$^{-1}$, respectively (hereafter ''luminosity'' means
the luminosity in the 3-100 keV energy band). These values are about two
orders of magnitude higher than the quiescent luminosity 
$\sim4\times10^{34}$ erg s$^{-1}$, see Fig.\ref{lcurve}). This formally moves
formally RX\,J0440.9+4431 from low-luminosity persistent Be/XRB class 
to transient one \citep{reig2011}.

In this paper, based on the data of Swift, RXTE and INTEGRAL observatories
in a wide energy band ($0.6-120$ keV), we report the discovery of a
cyclotron absorption line in the spectrum of
the X-ray pulsar RX\,J0440.9+4431 at the energy $\simeq32$ keV, that corresponds to a magnetic
field strength of $3.2\times10^{12} {\rm G}$ (Section 3). A comprehensive
analysis of variations of the source power density spectrum (PDS) with the
luminosity during type~I outbursts in April and September 2010 confirmed
this estimation (Section 4.1). Results of the timing analysis on
different time scales (pulse and orbital periods) are presented
in Sections 4.2 and 4.3.

\section{Observations and Data Analysis}

\begin{figure}
\includegraphics[width=\columnwidth,bb=20 170 570 705, clip]{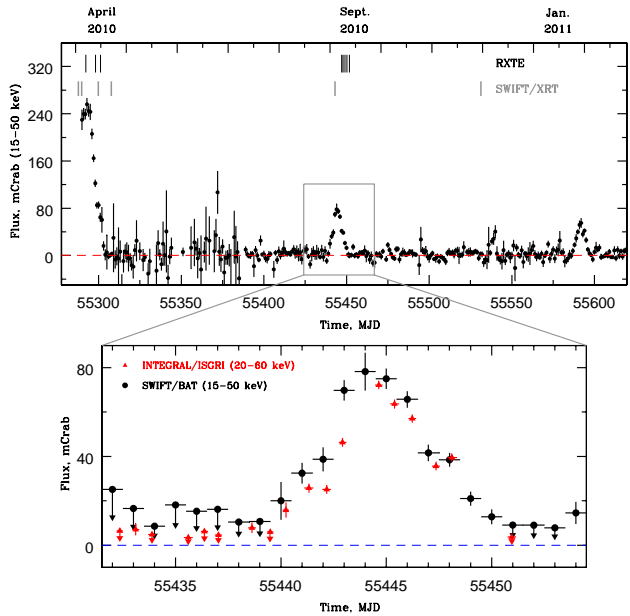}

\caption{One-day averaged light curve of the X-ray pulsar RX\,J0440.9+4431
obtained with the Swift/BAT monitor in the 15-50 keV energy band (top
panel). The thin upper marks correspond to the RXTE observations, the thick
lower marks -- to the Swift/XRT observations. The light curve obtained by
the INTEGRAL/IBIS telescope in the 20-60 keV energy band is shown in the
inset as a comparison to the Swift/BAT points. }\label{lcurve}

\end{figure}

To get the broadband view on the properties of the X-ray pulsar
RX\,J0440.9+4431 we used data from three currently operating observatories:
INTEGRAL, Swift and RXTE. The data collected during both outbursts in April
and September 2010 as well as in a quiescent state were analyzed. The source
light curve obtained with the Swift/BAT telescope in the $15-50$ keV energy
band is shown in Fig.\ref{lcurve} by circles. The INTEGRAL/ISGRI flux
measurements in the $20-60$ keV energy band are represented by triangles in
the inset.

Observations in the soft energy band $0.8-9$ keV were performed with the XRT
telescope onboard the Swift Observatory \citep{gehr2004} in different
intensity states (Obs.IDs 00031690, 00418109 and 00418178, thick lower
strokes in Fig.\ref{lcurve}) both in the Photon Counting (PC) and Window
Timing (WT) modes. As the observations in the PC mode performed in a
high-luminosity state were affected by the pile-up, in the subsequent
analysis we used only data obtained in the WT mode for uniformity. The
typical exposure during XRT observations was about of $3-5$ ksec each.

In the standard X-ray energy band ($4-20$ keV)\footnote{The latest
  improvement in the response matrix
  (http://www.universe.nasa.gov/xrays/programs/rxte/pca/doc/
  rmf/pcarmf-11.7/) allows formally to extend the PCA upper energy
  limit up to 50-60 keV. But our analysis showed that there are still
  significant systematic deviations at the energy of $\sim30$
  keV. Therefore, in this paper we limited oneself to using just a
  standard energy band.}  data from the RXTE/PCA spectrometer
\citep{br93} were used (Obs.ID 95418-01-XX-XX). The spectral analysis
was done using the Standard-2 data, temporal analysis -- using the
Good Xenon data with a high time resolution. Note that all RXTE
observations cover mostly the fading phase of both outbursts (thin
upper strokes in Fig.\ref{lcurve}) and have an exposure of several
kiloseconds.

This work also employs the INTEGRAL observatory \citep{win03} data acquired
during the Galactic latitude scan campaign at l=155$^o$ (Proposal ID
0720049, PI R. Krivonos). The source RX\,J0440.9+4431 was in the field of
view of both X-ray telescopes of the observatory: JEM-X (energy band $6-20$
keV) and IBIS/ISGRI (energy band $18-120$ keV). The reduction of the ISGRI
detector data was done using recently developed methods described by
\cite{kri2010b}. The source spectrum from the JEM-X telescope was extracted
with the standard OSA package version
9.0\footnote{http://isdc.unige.ch}. Both spectra (from JEM-X and IBIS
telescopes) were averaged for several days near the maximum of the September
2010 outburst with total exposures of $\sim 47$ and $\sim 180$ ksec,
respectively. Such a difference in the exposure time is connected with an
observational pattern and different fields of view of the instruments (see
\citet{win03} for details). The method used for ISGRI data spectra extraction
has a systematic error in the source flux determination of about
3\%. We included this systematic uncertainty in spectral analysis done with the
XSPEC package.

The final spectral and timing analysis for all instruments were done using
the standard tools of the FTOOLS/LHEASOFT 6.7 package.

\section{Spectral analysis}

\subsection{The Continuum Spectrum}

The spectrum of RX\,J0440.9+4431 in the $3-20$ keV energy band was obtained
by \cite{reig1999} using the RXTE observatory data acquired in 1998 in the
quiescent state (the corresponding luminosity in 3-30 keV energy band was
$\simeq3\times10^{34}$ erg s$^{-1}$). The source faintness didn't allow
those authors to distinguish meaningfully between the different spectral
models or determine unambiguously spectral parameters. Particularly, the
value of the photoelectric absorption was varied from its absence to
$\sim6\times10^{22}$ atom cm$^{-2}$ depending on the model, whereas the
interstellar absorption based on the measurements of the 21 cm emission in
the source direction is $\sim0.6\times10^{22}$ atom cm$^{-2}$
\citep{kalb2005}. Moreover, \citet{reig1999} also reported about a marginal
detection of a fluorescent iron line with an upper limit to its equivalent
width of $\sim100$ eV.

On the other hand proper knowledge of the continuum is crucial not only to
understand the physical mechanisms of the X-ray emission generation but also
detect possible line-like features in the source spectrum. Therefore, we
first investigated the spectrum of the continuum of RX\,J0440.9+4431.

\begin{figure}
\includegraphics[width=\columnwidth, bb=60 290 570 710, clip]{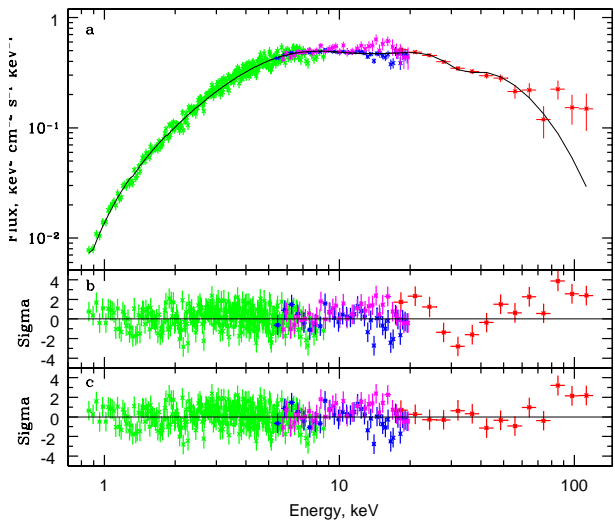}

\caption{Energy spectrum of RX\,J0440.9+4431 obtained during the outburst in
September 2010 with XRT (green points), JEM-X (magenta points), PCA (blue points)
and IBIS (red points) telescopes ({\it a}). The
residuals to the fit in units of sigma without ({\it b}) and with ({\it c})
a $\sim32$ keV cyclotron absorption line. }\label{specint}

\end{figure}

As mentioned above the XRT telescope observed RX\,J0440.9+4431 several times
both in the quiescent state and during the outbursts. In all cases,
independently of the spectral model used, the source spectrum didn't require
the inclusion of a strong photoelectric absorption. Its value varied
insignificantly in the range of $N_{\rm H}\simeq(0.2-0.5)\times10^{22}$ atom
cm$^{-2}$, which is compatible with the value of interstellar absorption in
the source direction. The quiescent spectrum was well fitted by a powerlaw
model with an exponential cutoff (\textit{cutoffpl} model in {\rm XSPEC}),
that is typical for X-ray pulsars, but the required cutoff energy was
unusually low ($\simeq1.8-2.4$ keV). The source spectra obtained in a high
state could not be described adequately by a single-component model. The
addition of a black-body component with a temperature of $kT_{\rm
BB}\simeq1.4-1.7$ keV to the powerlaw model significantly improved the fit
quality. As the detailed study of the spectral evolution of RX\,J0440.9+4431
is beyond the scope of this paper, we give here only as an example the
spectral parameters obtained during the high state in April 2010 (Obs.ID
00031690001, MJD 55299.9): temperature of the black-body component $kT_{\rm
BB}=1.50\pm0.05$ keV, photon index $\Gamma=1.13\pm0.27$, absorption $N_{\rm
H}=(0.46\pm0.07)\times10^{22}$ atom cm$^{-2}$.

Widening the considered energy band up to $0.6-20$ keV by the inclusion of
PCA/RXTE data doesn't change the described picture or spectral parameters
significantly; the only thing that was added is the exponential cutoff with
energies of $5-8$ keV. It is important to note that changes of the source
flux during outbursts by a factor of $\sim(2-3)$ didn't lead to changes of
the shape of the source spectrum, therefore we jointly fitted XRT and PCA
data just adding a normalization factor.

Finally, we could not detect the fluorescent iron line at 6.4 keV in any
luminosity state and obtained only an upper limit to its equivalent width
$\simeq50$ eV (90\% confidence level).

\begin{table}
\caption{Best-fit results in the energy range 0.6-120 keV}\label{tablspec}
\begin{tabular}{lll}
\hline
\hline
Parameter & Model I  & Model II  \\

\hline

$N_{\rm H}\times10^{22}$, cm$^{-2}$ & $0.47\pm0.04$ & $0.46\pm0.04$ \\
$kT_{\rm BB}$, keV                  & $1.57\pm0.03$ & $1.56\pm0.03$ \\
$R_{\rm BB}$, km                    & $0.91\pm0.04$ & $0.93\pm0.05$ \\
Photon index                        & $0.74\pm0.05$ & $0.75\pm0.05$ \\
$E_{cut}$, keV                      & $16.9\pm1.0$& $18.4\pm1.1$ \\
$\tau_{cycl}$                       &               & $0.37\pm0.06$ \\
$E_{cycl}$, keV                     &               & $31.9\pm1.3$  \\
$\sigma_{cycl}$, keV                &               & $6$ (fixed)   \\
$RF_{JEM-X}$                        & $1.08\pm0.02$ & $1.08\pm0.02$ \\
$RF_{PCA}$                          & $0.55\pm0.02$ & $0.55\pm0.02$ \\
$RF_{IBIS}$                         & $1.16\pm0.06$ & $1.26\pm0.06$ \\
$\chi^2$ (d.o.f)                    & $1.315 (331)$  & $1.223 (329)$  \\
\hline\\
\end{tabular}\\
All uncertainties correspond to one standard deviation  \\
\end{table}

\subsection{Detection of a Cyclotron Absorption Line}

The cyclotron resonance scattering features (CRSFs) with typical energies of
$\simeq20-50$ keV are observed at the moment in more than a dozen X-ray
pulsars (see, e.g., \cite{cob2002}; \cite{fil2005}). These features are
thought to be due to photon resonant scattering by electrons at Landau
orbits in the strong magnetic field (of the order of 10$^{12}$ G) of the
neutron star and, thus, their registration in the source spectrum gives a
direct measurement of the magnetic field value.

The pulsar RX\,J0440.9+4431 was detected for the first time in the
hard X-ray band ($>20$ keV) by the IBIS telescope during the September
2010 outburst. Since the source flux was relatively low, only the
spectrum averaged over the whole outburst had enough statistics for
the following analysis. To reconstruct the broadband ($0.6-120$ keV)
spectrum we used data from XRT (Obs.ID 00031690003, MJD 55443.03),
JEM-X and IBIS (MJD 55440-55449) telescopes; an averaged flux in the
3-100 keV energy band during these sets of observations was
$\sim2.3\times10^{-9}$ erg cm$^{-2}$ s$^{-1}$. To get more strict
constraints on the spectrum parameters we introduced to the fitting
procedure also PCA data (Obs.ID 95418-01-03-00, MJD 55447.96). It is
necessary to note, that this observation was performed several days
after the outburst maximum, when the source flux was several times
less. But, as it was mentioned in \S 3.1, the spectrum shape remains
practically the same with changes of the source flux, therefore we can
jointly fitted all these data leaving the normalization factor free.

The energy spectrum of RX\,J0440.9+4431 is shown in Fig.\ref{specint}a
(green points -- XRT data, magenta points -- JEM-X data, blue points
-- PCA data and red points -- IBIS data). Based on the previous
analysis (\S3.1) it was approximated by the two-component model
(black-body radiation and powerlaw with an exponential cutoff) with a
photoelectric absorption (Model I). Best-fit parameters are listed in
Table \ref{tablspec} ($RF_{JEM-X}$, $RF_{PCA}$ and $RF_{IBIS}$ are
renormalization factors between different instruments, $RF_{XRT}=1.0$,
was fixed). Note, that a difference between the obtained photon index
and the cutoff energy from the best-fit results in the standard energy
band is due to the inclusion of higher energy data.  This model
describes the data relatively well everywhere except around $\sim30$
keV, where one can clearly see an absorption like deviation from the
continuum model (Fig.\ref{specint}, middle panel). It is worth noting
that the appearance of this feature is not a sequence of the model of
the continuum used. We investigated several other combinations of
spectral components to describe the continuum (NPEX\,+\,black-body,
{\em highecut}\,+\,black-body) and found that: 1) all of them
approximate it as well as {\em cutoffpl}\,+\,black-body, but have more
parameters; 2) residuals for all of them demonstrate a prominent
absorption-like feature near 30 keV. We also checked that this feature
is not an artifact connected with the response matrix of the ISGRI
detector. For this purpose spectra of a number of sources with a
similar intensity and without spectral features in the range of
interest were reconstructed from nearby observations of the Galactic
Bulge.  No unusual absorption features were found in the spectra of
these sources. Thus, we can conclude that the absorption feature in
the spectrum of RX\,J0440.9+4431 is the attribute of the source
arising due to a cyclotron resonant scattering.

To describe the detected feature three different spectral models were used:

1. multiplicative XSPEC model {\it cyclabs} in the form of:
\begin{equation}
$$\exp\left(\frac{-\tau_\mathrm{cycl}(E/E_\mathrm{cycl})^2\sigma_\mathrm{cycl}^2}
{(E-E_\mathrm{cycl})^2+\sigma_\mathrm{cycl}^2}\right),$$
\end{equation}

\noindent where $E_\mathrm{cycl}$, $\sigma_\mathrm{cycl}$, and $\tau_\mathrm{cycl}$ are the
line central energy, width, and depth, respectively  (\cite{mih90});

2. multiplicative XSPEC model {\it gabs} in the form of:
\begin{equation}
$$\exp\left(\left(\frac{-\tau_\mathrm{cycl}}{\sqrt{2\pi}\sigma_\mathrm{cycl}}\right)
\exp\left(\frac{-(E-E_\mathrm{cycl})^2}{2\sigma_\mathrm{cycl}^2}\right)\right)$$
\end{equation}

\noindent (see eq. 6 and 7 in \cite{cob2002});

3. simple additive Gaussian with a negative normalization.

Best-fit parameters for the case of the {\it cyclabs} model (Model II)
are summarized in Table \ref{tablspec}. It is seen that an addition of
the cyclotron absorption line leads to a substantial reduction by
$\chi^2 \simeq 30$ for 2 d.o.f. Nevertheless, to estimate the
significance of the line detection we performed $10^5$ Monte-Carlo
simulations of the spectra based on the best-fit model without a
line. Then we fitted the obtained spectra adding the {\it cyclabs}
component and analyzed the distribution of the line depth (a similar
approach was used by \cite{rea2005}). Parameters of the model for the
simulated spectra were the same as for the real spectrum: the line
width was fixed at 6 keV, but the line depth and energy were free
parameters. None of the absorption lines detected in the simulated
spectra was deep as in the observed one. Thus, based on these
simulations we can conclude that the significance of the cyclotron
absorption line in the spectrum of RX\,J0440.9+4431 is higher than
$\sim4\sigma$.

The best-fit positions of the line energy in other cases depend on the model used, but
nevertheless agree with the {\it cyclabs} results within error bars:
$E_{cycl}=34.2\pm1.5$ keV for the {\it gabs} model and $E_{cycl}=30.6\pm1.6$
keV for the Gaussian one. It is important to note that in all three models
we could not correctly constrain the line width, probably due to a source
faintness and small depth of the line, therefore it was fixed at the value
of 6 keV, that is typical for cyclotron lines in the spectra of X-ray
pulsars (see, e.g., \cite{cob2002}, \cite{fil2005}).

From the measured value of the cyclotron line energy we can estimate the
magnetic field on the neutron star surface

$$B_{NS}=\frac{1}{\sqrt{\left(1-\frac{2GM_{NS}}{R_{NS}c^2}\right)}} \frac{E}{11.6}\simeq3.2\times10^{12} {\rm G}$$

\noindent where $R_{NS}=15$ km \citep{sul2011} and $M_{NS}=1.4M_{\sun}$ --
are the neutron star radius and mass estimates, respectively.

Finally, it is necessary to note that despite the significant improvement of
the fit quality an excess of emission at energies higher than $\sim$100 keV
still remains. Formally it can be fitted by an additional powerlaw component,
but due to the lack of statistics at such high energies it is impossible to
obtain significant restrictions to its parameters and to make conclusions
about its nature.

\section{Timing analysis}

\subsection{Power Spectra}

\begin{figure}
\vbox{
\includegraphics[width=\columnwidth,bb=50 275 550 680,clip]
{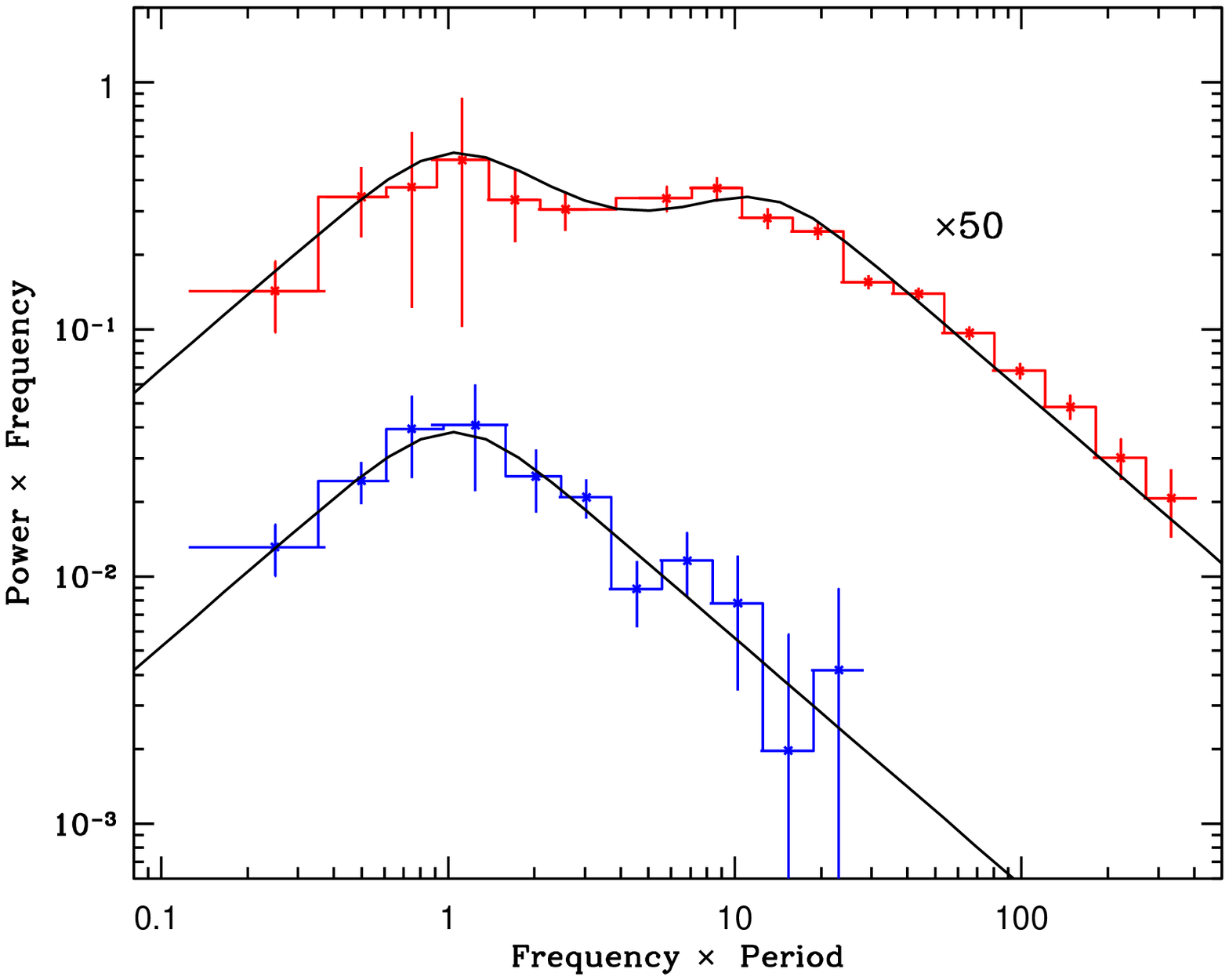}
\includegraphics[width=\columnwidth,bb=65 430 575 690,clip]
{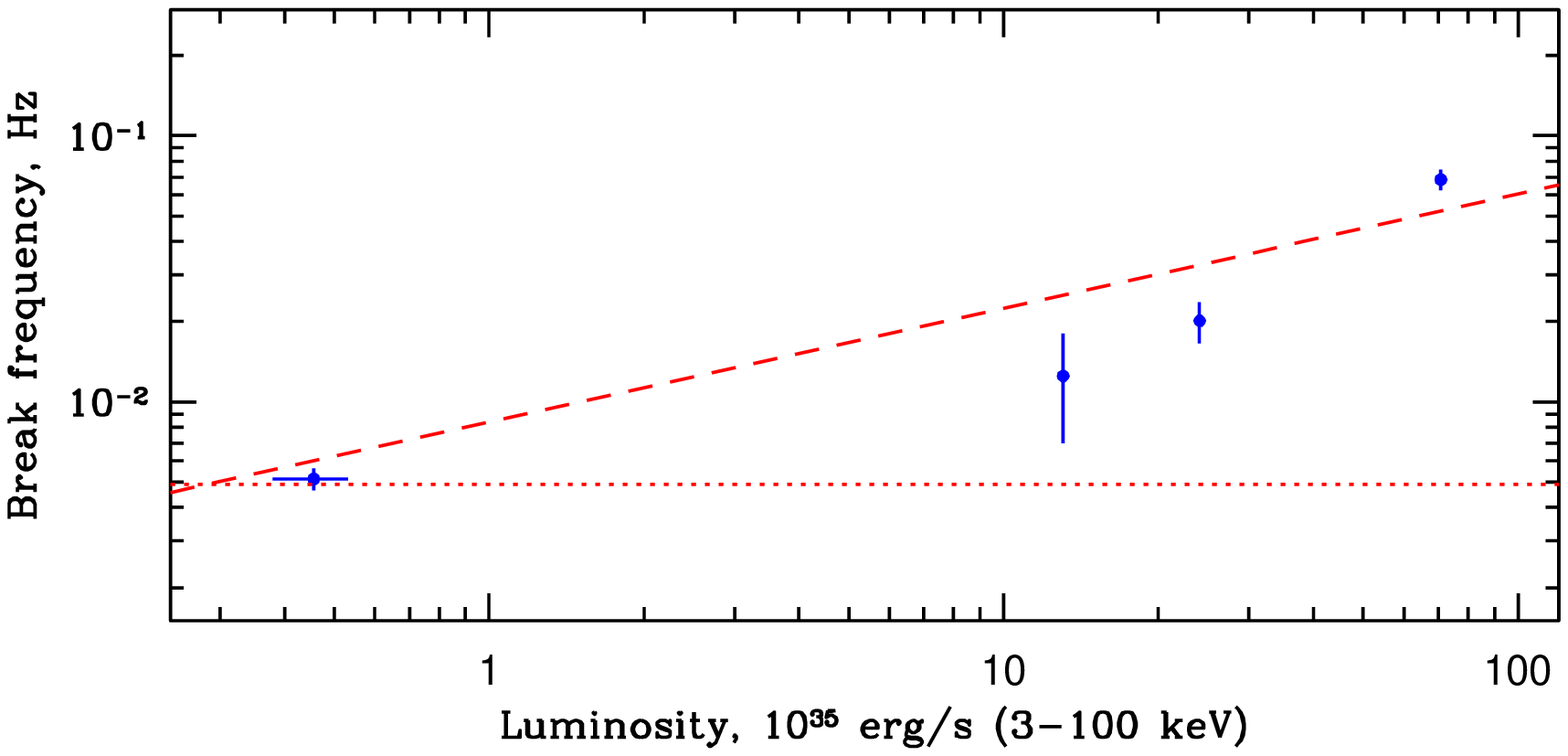}}

\caption{{\it Top:} Power density spectra of RX\,J0440.9+4431 in different
luminosity states. The spectrum in the bright state is multiplied by a
factor of 50 for clarity.  The frequency shown along the X-axis is expressed
in units of the compact object spin frequency. The power density spectra are
multiplied by the frequency. The solid lines show the best-fit models (see text
for details). {\it Bottom:} dependence of the break frequency in the noise
power spectrum of RX\,J0440.9+4431 on the bolometric luminosity of the
source.  The dashed line shows the prediction of the model described in the
text. The dotted line shows the neutron star spin frequency.}\label{pspec}

\end{figure}

Accreting X-ray pulsars are characterized by a specific shape of the noise
power spectrum. Due to the presence of the strong magnetic field of the
neutron star, the disk-like accretion flow is disrupted at the
magnetospheric boundary. This divides the flow into two distinct parts --
the accretion disk and the magnetospheric flow, which have different noise
properties \citep{revn09}.

It was shown by \cite{revn09} that the break frequency in the noise power
spectra of accreting X-ray pulsars reflects the timescale of noise
generation at the inner boundary of the accretion disk/flow, and its value
depends on the mass accretion rate in the binary system. An increase of the
mass accretion rate reduces the size of the magnetosphere (and hence the
inner radius of the disk), so that the characteristic frequency at the inner
edge of the disk/flow is increased. This property of the power spectrum can
be used for estimations of the magnetic moment of the accreting compact
star.

Due to the source faintness and short duration of the observations it was
difficult to build the qualitative power spectrum for each RXTE
observation. But we succeeded by averaging them into four points with
mean luminosity $\sim4\times10^{34}$ erg s$^{-1}$ (ID 30111-01-01-00),
$\sim1.3\times10^{35}$ erg s$^{-1}$ (IDs 95418-01-04-00, 95418-01-04-01),
$\sim2.4\times10^{35}$ erg s$^{-1}$ (IDs 95418-01-02-01, 95418-01-03-00,
95418-01-03-01) and $\sim7.1\times10^{36}$ erg s$^{-1}$ (IDs 95418-01-01-00,
95418-01-02-00) (here we took into account the fact that the spectrum
shape is nearly constant with the luminosity and recalculated the PCA flux
in the 3-20 keV energy band to the 3-100 keV one). In Fig.\ref{pspec} (top
panel) two power spectra of RX\,J0440.9+4431 are shown for different
luminosities. The blue histogram corresponds to the PCA observation
30111-01-01-00 (MJD 50843.27) in a low luminosity state with
$L_X\simeq4\times10^{34}$ erg s$^{-1}$, the red histogram represents the
bright state with an averaged luminosity of $L_X\sim7.1\times10^{36}$ erg
s$^{-1}$ (multiplied by a factor of 50 for clarity). The frequency
shown along the X-axis is expressed in units of the compact object spin
frequency: the pulse period was 202.5 \citep{reig1999} and 205.0 s (this
work) in the low and high luminosity states, respectively; see Sect. 4.2 for
details. \cite{revn09} showed that the noise power spectra of all examined
sources have a similar power-law slope ($P \sim f^{-2}$) above the break
frequency, irrespective of the power spectra form at lower frequencies. To
fit the PDSs of X-ray pulsars these authors proposed a simple model in the form of
$P\propto f\times(1+(f/f_b)^4)^{-0.5}$.

Solid lines in Fig.\ref{pspec} represent results of the approximation of the
power spectra of RX\,J0440.9+4431 with the described model: in the
low-intensity state one-component model was used; in the high-intensity
state the second component in the same form was added to the model. Thereby
an additional noise component was taken into account. This component is
presumably generated in the ring of the accretion disk between the radii
corresponding to the size of the magnetosphere at the high accretion rate
(small radius) and low accretion rate (large radius) \citep{revn09}.

Taking the expression for the frequency of the Keplerian rotation as
$2\pi\nu_{K}=(GM)^{1/2}R_{M}^{-3/2}$ and for magnetospheric radius as
$R_{M}=\mu^{4/7}(2GM)^{-1/7}\dot{M}^{-2/7}$ \citep{bild97} one can show that
the break frequency then depends on the neutron star's magnetic moment and
mass accretion rate as:

\begin{equation}\label{aqfb}
$$f_b = \frac{2^{3/14}}{2\pi}(GM)^{5/7}\mu^{-6/7}\dot{M}^{3/7},$$
\end{equation}

\noindent where $f_{b}$ is the break frequency, $\mu$ is the dipole magnetic
moment of the neutron star and $\dot{M}$ is the mass accretion rate.

Substituting in Eq.\ref{aqfb} $R_{NS}=15$ km, $M_{NS}=1.4M_{\sun}$, 3.3 kpc
for the distance to the source and $L_x=0.1c^2\dot{M}$ we can approximate
the measured values of the break frequency (dashed line in the bottom panel
of Fig.\ref{pspec}) and determine the strength of the magnetic field of the
neutron star as $B=3.3\times10^{12}$ G, which is in very good agreement with
the one derived from the cyclotron absorption line energy detected in the
source spectrum.

\begin{figure}
\vbox{
\includegraphics[width=\columnwidth,bb=165 140 500 710,clip]
{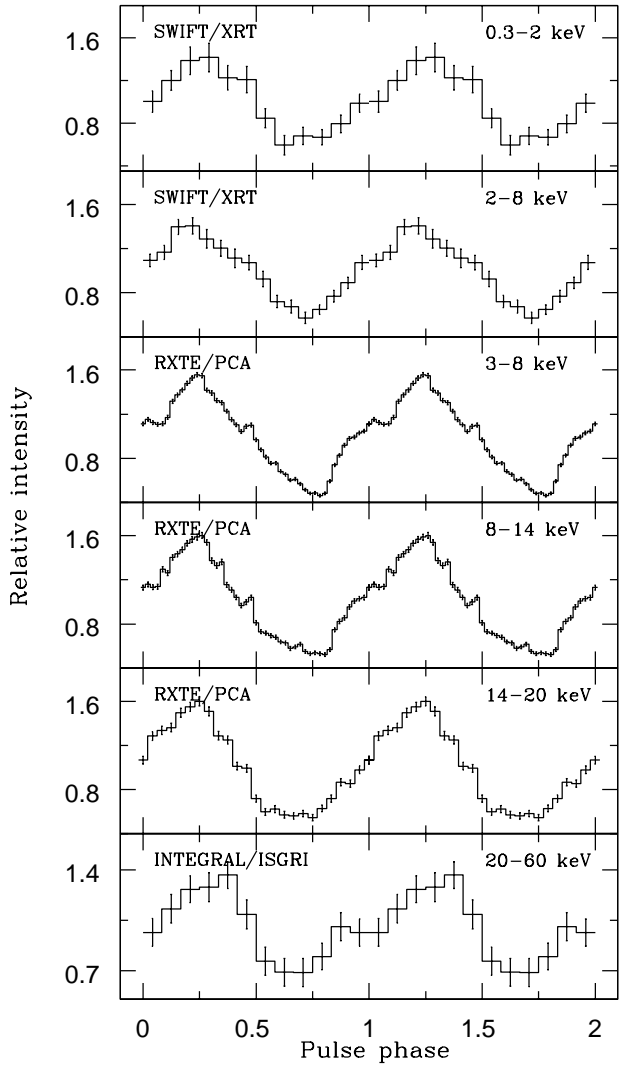}}

\caption{Pulse profiles of RX\,J0440.9+4431 in different energy
bands (normalized by a mean flux). Data from Swift/XRT (observation 00031690003, MJD 55443.03),
RXTE/PCA (observation 95418-01-04-00, MJD 55449.04) and INTEGRAL/ISGRI (revolutions 963, 964, 965; MJD 55441.1-55448.4)
were used (an average luminosity is about $1.5\times10^{36}$ erg s$^{-1}$).
The profile is shown twice for a clarity.}\label{ppener}
\end{figure}
\begin{figure}
\vbox{
\includegraphics[width=\columnwidth,bb=165 220 500 710,clip]
{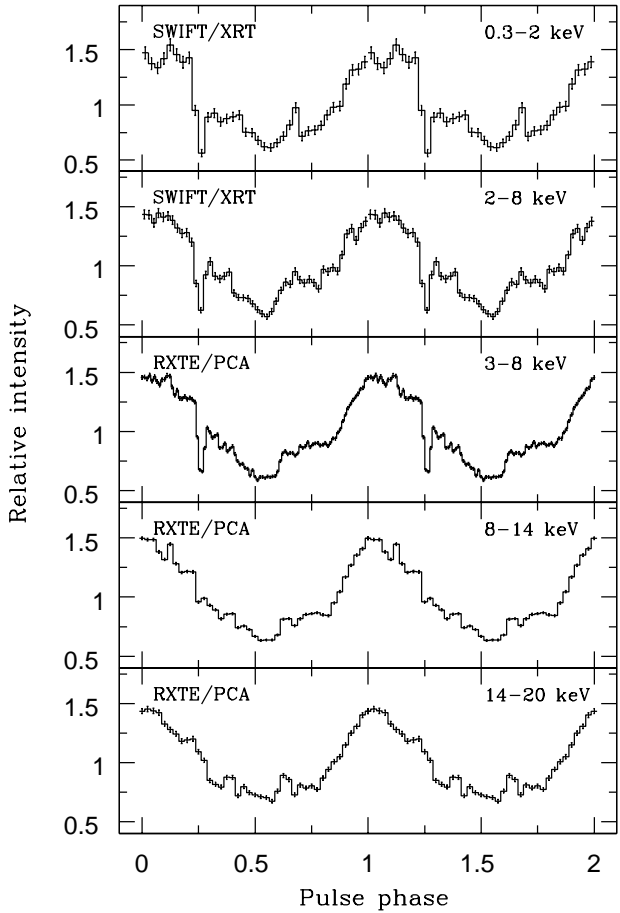}}

\caption{The same as in Fig.\ref{ppener}, but for the high-luminosity state with $L_X\simeq8\times10^{36}$ erg s$^{-1}$. Data from Swift/XRT (observation 00418178000, MJD 55289.85) and RXTE/PCA (observation 95418-01-01-00, MJD 55292.33) were used. The profile is shown twice for a clarity.}\label{ppener1}
\end{figure}

\subsection{Pulse Profile}

It is well known that the pulse profiles of X-ray pulsars are strongly
dependent on photon energy and luminosity (see, e.g., \citet{lut09} and
references therein). This is especially prominent in a soft energy band
strongly affected by the absorption in the vicinity of the neutron
star. Therefore the evolution of the pulse profile can provide an important
information for understanding the properties and spatial distribution of
matter around the neutron star. To reconstruct pulse profiles we determined
the pulse period during each observation. We didn't detect any significant
variations of the pulse period during the outbursts therefore in the
following analysis we used the value averaged over the whole September 2010
outburst $205.0\pm0.1$ s. This period is significantly longer than the one
measured in 1998 $\simeq202.5$ s \citep{reig1999}, that indicates the
deceleration of the neutron star rotation over the last decade with an
average rate of $\dot P/P \simeq 10^{-3}$ yr$^{-1}$.

In Fig.\ref{ppener} the pulse profile of RX\,J0440.9+4431 in different
energy bands is shown to illustrate the evolution of it's shape with the
energy in the low luminosity state ($L_X\sim1.5\times10^{36}$ erg
s$^{-1}$). In general the pulse profile of RX\,J0440.9+4431 has a
sinusoidally-like single-peaked shape in a wide energy band. But on the
light curves with high statistics (obtained with the PCA spectrometer in
$3-8$ and $8-14$ keV energy bands) some substructures become
apparent. Particularly, one can see at least two prominent features:
subpulses before and after the main peak.

In the high luminosity state (the mean luminosity is
$L_X\simeq8\times10^{36}$ erg s$^{-1}$) the pulse profile has approximately
the same shape, but shows another feature -- a dip-like structure at a phase
of $\sim0.25$ after the main peak. This feature is clearly seen at the
energies below 8 keV (Fig.\ref{ppener1}). \cite{usui2011} interpreted this
feature as a partial eclipse of the emission region by an accretion column
of the neutron star. It is interesting to note that in the low luminosity
state (Fig.\ref{ppener}) such an absorption feature wasn't detected in the
pulse profile. Probably its appearance can be connected with changes of the
geometry of the accretion column or with changes of the neutron star
orientation relative to the observer.

\begin{figure}
\vbox{
\includegraphics[width=\columnwidth,bb=25 275 515 675,clip]
{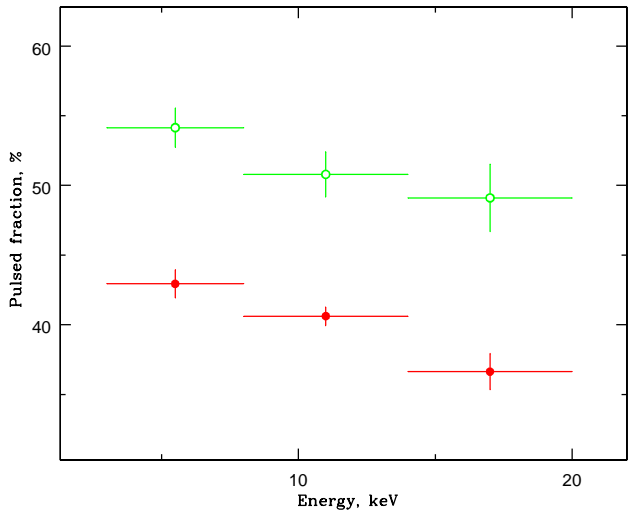}}

\caption{Pulsed fraction dependence on the energy for two intensity states:
$L_X\simeq1.3\times10^{36}$ erg s$^{-1}$ (open circles, RXTE observation
95418-01-04-00, MJD 55449.04) and $L_X\simeq8\times10^{36}$ erg s$^{-1}$
(filled circles, an RXTE observation 95418-01-01-00, MJD
55292.33).}\label{pfrac}

\end{figure}

The pulsed fraction\footnote{It was determined as
  $\mathrm{PF}=(I_\mathrm{max}-I_\mathrm{min})/(I_\mathrm{max}+I_\mathrm{min})$,
  where $I_\mathrm{max}$ and $I_\mathrm{min}$ are maximum and minimum
  intensities in the pulse profile of an X-ray pulsar, respectively.}
of the source emission in the standard X-ray band ($4-20$ keV) is
relatively high: (40.9$\pm$0.1)\% at high luminosities
($\sim8\times10^{36}$ erg s$^{-1}$, MJD 55292.33) and increases with
the fading of the pulsar intensity. In particular, in observations
95418-01-04-00 (MJD 55449.04) and 95418-01-04-01 (MJD 55450.23), when
the source luminosity dropped to $\sim1.3\times10^{36}$ erg s$^{-1}$
and $\sim9\times10^{35}$ erg s$^{-1}$, the pulsed fraction came up to
the values of (50.1$\pm$0.1)\% and (61.5$\pm$1.0)\%,
respectively. Such a behavior is typical for X-ray pulsars
\citep{lut09}. However for RX\,J0440.9+4431 the pulsed fraction shows
a tendency to decrease towards higher energies (Fig.\ref{pfrac}).

\subsection{Orbital Period}

In 2010-2011 three outbursts from RX\,J0440.9+4431 were sequentially
detected by the Swift observatory (see Fig.\ref{lcurve}). The peak
luminosities of all three were relatively low: $\sim9\times10^{36}$,
$\sim3\times10^{36}$ and $\sim2\times10^{36}$ erg s$^{-1}$ (assuming
distance to the source of 3.3 kpc), that allows us to consider them as
Type~I outbursts. It was mentioned above that such events are
separated by one orbital period of the binary system. Taking into
account that all three outbursts were roughly equally spaced in time
we estimated the orbital period of RX\,J0440.9+4431 as $\sim155$ days
(a mean time between starts of the outbursts). This value is well
agreed with the period of about 150 days derived from the Corbet
diagram of $P_{spin}$ versus $P_{orbit}$ \citep{corb1986}. An accurate
determination of the orbital parameters of the binary system should be
based on measurements of the Doppler delay of the pulse period due to
the orbital motion of the neutron star in the binary system and
requires much longer observations in a combination with the high
sensitivity and timing accuracy.

\section{Conclusion}

In this work we have presented the temporal and spectral analysis of the
poorly studied X-ray pulsar RX\,J0440.9+4431 with the Be star
companion. Using the data of Swift, RXTE and INTEGRAL Observatories the
broadband spectrum of the source was reconstructed for the first time. It
was shown that the spectrum can be well approximated by the two-component
model, including the black-body emission with the temperature of $kT_{\rm
BB}\simeq1.5$ keV and the high-energy component in the form of a powerlaw
with an exponential cutoff. Such a spectral shape is similar to that
observed for other low-luminosity X-ray pulsars in Be binary systems, like X
Persei or RX\,J0146.9+6121, and reminiscent of the thermal emission from the
neutron star polar caps and the comptonized nonthermal emission at high
energies (see, e.g., \citet{disalvo98,lapal06} and references therein). The
corresponding black-body emitting radius $R_{BB}$ (see Table \ref{tablspec})
is about 1.1 km which agrees with the expected radius of an accretion column
of the neutron star $R_{col}\sim0.1R_{NS}$ (see, e.g., \cite{bs76}).

We have discovered a prominent absorption feature at the energy $\sim32$ keV
in the spectrum of the source. Interpreting it as a cyclotron resonance
scattering feature the magnetic field strength of the neutron star was
estimated as $B\simeq3.2\times10^{12}$ G. The obtained value is in very
close agreement with one derived from the properties of the noise power
spectra.

The pulse profile of RX\,J0440.9+4431 has a sinusoidally-like single-peaked
shape, which is stable in relation to the changes of the source luminosity
and energy band. Nevertheless in the high intensity state we found a
dip-like structure at the pulse phase of $\sim0.25$ after the main peak that
can be interpreted as a partial eclipse of the emission region by the
accretion column of the neutron star. The pulsed fraction of the source
emission is growing with the decrease of the source luminosity, that is
typical for X-ray pulsars \citep{lut09}. On the other hand, in contrast
to other X-ray pulsars, for RX\,J0440.9+4431 the pulsed fraction shows
a tendency to decrease towards higher energies.

Based on the recurrence time between Type I outbursts the orbital period of
$\sim155$ days was supposed for the binary system. The gradual decrease of
the total energy release during these transient events can be probably
connected with the existence of the circumstellar disk around the Be star
and its variability in time (see, e.g., \citet{clark03,riv01}).

\section*{Acknowledgments}

Authors thank Mike Revnivtsev and Valery Suleimanov for helpful and useful
discussions. This work was supported by the program
``Origin, Structure and Evolution of the Objects in the Universe'' by the
Presidium of the Russian Academy of Sciences, grant no.NSh-5069.2010.2 from
the President of Russia, Russian Foundation for Basic Research (grants
11-02-01328 and 11-02-12285-ofi-m-2011), State contract
14.740.11.0611 and the Academy of Finland grant 127512.
The research used the data obtained from the HEASARC Online
Service provided by the NASA/Goddard Space Flight Center, European and
Russian INTEGRAL Science Data Centers. The results of this work are
partially based on observations of the INTEGRAL observatory, an ESA project
with the participation of Denmark, France, Germany, Italy, Switzerland,
Spain, the Czech Republic, Poland, Russia and the United States.

\label{lastpage}


\begin{thebibliography}{99}

\bibitem[\protect\citeauthoryear{Basko \& Sunyaev}{1976}]{bs76}
Basko M.M., Sunyaev R.A., 1976, MNRAS, 175, 395

\bibitem[\protect\citeauthoryear{Bildsten et al.}{1997}]{bild97}
Bildsten L., Chakrabarty D., Chiu J., et al., 1997, ApJS, 113, 367

\bibitem[\protect\citeauthoryear{Bradt et al.}{1993}]{br93}
Bradt H.V., Rothschild R.E., Swank J.H., 1993, A\&AS, 97, 355

\bibitem[\protect\citeauthoryear{Clark et al.}{2003}]{clark03}
Clark J., Tarasov A., Panko E., 2003, A\&A, 403, 239

\bibitem[\protect\citeauthoryear{Coburn et al.}{2002}]{cob2002}
Coburn W., Heindl W. A., Rothschild R. E., Gruber D. E., Kreykenbohm I., Wilms J.,
Kretschmar P., Staubert R., 2002, ApJ, 580, 394

\bibitem[\protect\citeauthoryear{Corbet}{1986}]{corb1986}
Corbet R.H.D., 1986, MNRAS, 220, 1047

\bibitem[\protect\citeauthoryear{di Salvo et al.}{1998}]{disalvo98}
di Salvo T., Burderi L., Robba N., Guainazzi M., 1998, ApJ, 509, 897

\bibitem[\protect\citeauthoryear{Filippova et al.}{2005}]{fil2005}
Filippova E., Tsygankov S., Lutovinov A., Sunyaev R., 2005, Astron. Lett., 31, 729

\bibitem[\protect\citeauthoryear{Gehrels et al.}{2004}]{gehr2004}
Gehrels N., Chincarini G., Giommi P. et al., 2004, ApJ, 611, 1005

\bibitem[\protect\citeauthoryear{Kalberla et al.}{2005}]{kalb2005}
Kalberla P., Burton W., Hartmann D, Arnal E., Bajaja E., Morras R., Poeppel W., 2005, A\&A, 440, 775

\bibitem[\protect\citeauthoryear{Krivonos et al.}{2010a}]{kri2010a}
Krivonos R., Tsygankov S., Lutovinov A., Turler M., Bozzo E., 2010, Astron. Telegram, 2828

\bibitem[\protect\citeauthoryear{Krivonos et al.}{2010b}]{kri2010b}
Krivonos R., Revnivtsev M., Tsygankov S., Sazonov S., Vikhlinin A., Pavlinsky M.,
Churazov E., Sunyaev R., 2010, A\&A, 519, A107

\bibitem[\protect\citeauthoryear{La Palombara \& Mereghetti}{2006}]{lapal06}
La Palombara N., Mereghetti S., 2006, A\&A, 455, 283

\bibitem[\protect\citeauthoryear{Lutovinov \& Tsygankov}{2009}]{lut09}
Lutovinov A.A., Tsygankov S.S., 2009, Astron. Lett., 35, 433

\bibitem[\protect\citeauthoryear{Mihara et al.}{1990}]{mih90}
Mihara T., Makishima K., Ohashi T., et al., 1990, Nature, 346, 250

\bibitem[\protect\citeauthoryear{Morii et al.}{2010}]{morii2010}
Morii M., Kawai N., Sugimori K. et al., 2010, Astron. Telegram, 2527

\bibitem[\protect\citeauthoryear{Motch et al.}{1997}]{motch97}
Motch C., Haberl F., Dennerl K., Pakull M., Janot-Pacheco E., 1997, A\&A, 323, 853

\bibitem[\protect\citeauthoryear{Rea et al.}{2005}]{rea2005}
Rea N., Oosterbroek T., Zane S., Turolla R., Mendez M., Israel G. L.,
Stella L., Haberl F., 2005, MNRAS, 361, 710

\bibitem[\protect\citeauthoryear{Reig \& Roche}{1999}]{reig1999}
Reig P., Roche P., 1999, MNRAS, 306, 100

\bibitem[\protect\citeauthoryear{Reig et al.}{2005}]{reig2005}
Reig P., Negueruela I., Fabregat J., Chato R., Coe M. J., 2005, A\&A, 440, 1079

\bibitem[\protect\citeauthoryear{Reig}{2011}]{reig2011}
Reig P., 2011, Ap\&SS, 332, 1

\bibitem[\protect\citeauthoryear{Revnivtsev et al.}{2009}]{revn09}
Revnivtsev M., Churazov E., Postnov K., Tsygankov S., 2009, A\&A, 507, 1211

\bibitem[\protect\citeauthoryear{Rivinius et al.}{2001}]{riv01}
Rivinius Th., Baade D., Stefl S., Maintz M., 2001, A\&A, 379, 257

\bibitem[\protect\citeauthoryear{Suleimanov et al.}{2011}]{sul2011}
Suleimanov V., Poutanen J., Revnivtsev M., Werner K., 2011, submitted, arXiv:1004.4871

\bibitem[\protect\citeauthoryear{Tsygankov et al.}{2011}]{tsy2011}
Tsygankov S., Lutovinov A., Krivonos R., 2011, Astron. Telegram, 3137

\bibitem[\protect\citeauthoryear{Usui et al.}{in preparation}]{usui2011}
Usui R. et al., in preparation

\bibitem[\protect\citeauthoryear{White et al.}{1983}]{wh83}
White N., Swank J., Holt S., 1983, ApJ, 270, 711

\bibitem[\protect\citeauthoryear{Winkler et al.}{2003}]{win03}
Winkler C., Courvoisier T.J.-L., Di Cocco G., et al.,  A\&A, 411, L1



\end{thebibliography}
\end{document}